\documentclass[aps,twocolumn,pra,superscriptaddress,amsmath,showpacs,tightenlines]{revtex4-1}
\usepackage{amssymb}
\usepackage{amsmath}
\usepackage{graphicx}
\usepackage{subfigure}
\usepackage{natbib}
\usepackage{epsfig}
\usepackage{amsfonts}
\usepackage{mathrsfs}
\usepackage{CJK}
\usepackage{xcolor}
\usepackage{threeparttable}
\usepackage[toc,page,title,titletoc,header]{appendix}

\begin{document}

\title{Phase-modulated {Autler-Townes splitting} in a giant-atom system within waveguide QED}

\author{Wei Zhao}
\affiliation{Center for Quantum Sciences and School of Physics, Northeast Normal University, Changchun 130024, China}
\author{Yan Zhang}
\affiliation{Center for Quantum Sciences and School of Physics, Northeast Normal University, Changchun 130024, China}
\author{Zhihai Wang}
\email{wangzh761@nenu.edu.cn}
\affiliation{Center for Quantum Sciences and School of Physics, Northeast Normal University, Changchun 130024, China}

\begin{abstract}
The nonlocal emitter-waveguide coupling, which gives birth to the so called giant atom, represents a new paradigm in the field of quantum optics and waveguide QED. In this paper, we investigate the single-photon scattering in a one-dimensional waveguide on a two-level or three-level giant atom. Thanks to the natural interference induced by the back and forth photon transmitted/reflected between the atom-waveguide coupling points, the photon transmission can be dynamically controlled by the periodic phase modulation via adjusting the size of the giant atom. For the two-level giant-atom setup, we demonstrate the energy shift which is dependent on the atomic size. {For the driven three-level giant-atom setup, it is of great interest that, the Autler-Townes splitting is dramatically modulated by the giant atom, in which the width of the transmission valleys (reflection range) is tunable in terms of the atomic size.} Our investigation will be beneficial to the photon or phonon control in quantum network based on mesoscopical or even macroscopical quantum nodes involving the giant atom.
\end{abstract}


\maketitle
\section{INTRODUCTION}
In the field of quantum optics, the study of the interaction between light and matter is one of the long-lived subjects. Recently, the light-matter interaction in waveguide structures has attracted much attention, which leads {to lots} of theoretical and experimental {works} in waveguide QED community~\cite{xg2017,cj2021,dr2017,ljy2020,gz2017,gz2018,gz2019,ii2020}{, such as dressed or bound states~\cite{hz2010,cj2019,ts2016,gc2016,es2017,pt2017,gc2019}, phase transitions~\cite{qj2011,mf2017,lq2019}, single-photon devices~\cite{jt2005,de2007,lz2008}, exotic topological and chiral phenomena~\cite{mr2014,vy2014,cg2016,im2016,sm2019,mb2019}, where the wavelength of light (or microwave field) is usually tens or hundreds of times larger than the size of the natural/artificial atoms constituting} the matter~\cite{pg1983,dl2003,aw2004,rm2005,sh2013}.
Therefore, the light-matter interaction is usually modelled by the dipole approximation, where the atoms are regarded as point-like dipoles~\cite{dw2018}.

{However, in recent years, the artificial superconducting transmon qubit coupled by the acoustic waves~\cite{sd1986,dm2007,rm2017}, of which the size is comparable to the wavelength of the phonons, is named as ``giant atom" and has been successfully realized in experiments~\cite{mv2014} .}
Alternatively, the giant-atom model can also be realized in the superconducting transmission line setup, where the capacitive or inductive coupling allows more than one coupling points between the microwave field and the qubit~\cite{bk2020,AM2020}.
Moreover, {using the cold atomic system, a theoretical scheme for the realization of giant atom has been proposed} in dynamical state-dependent optical lattices~\cite{AG2019}. In the giant-atom community, a lot of new phenomena {not existing} in the conventional small {atomic} system have been predicted, such as frequency dependent relaxation~\cite{af2014}, non-exponential decay~\cite{lg2017,ga2019,sg2020}, tunable bound state~\cite{lg2020,xw2020} as well as decoherence free subspace~\cite{af2018,ac2020} (For a recent review, see Ref.~\cite{review}).
The underlying physics behind these phenomena is {the} interference and retarded effect during the photon/phonon propagating process between the different coupling points.

On the other hand, {the dynamical control of the single-photon transmission} is a hot topic in constructing quantum networks~\cite{hj2008,sr2012,pl2017}, motivated by the photon based quantum information processing.
Photons provide a reliable transmission of quantum information and the waveguide is often seen as photonic channels in quantum network, with atoms (or artificial atoms) acting as quantum nodes.  Along this line, people have proposed lots of schemes to realize single-photon device, in which the propagating of the photon in channels is controlled on demand by adjusting the nature of the quantum nodes~\cite{jt2005,de2007,lz2008,jt2009,pl2010,wb2013,zh2014,wz2017}. {Furthermore, for a three-level quantum node, the electromagnetically induced transparency (EIT)~\cite{AA2010} and Autler-Townes splitting (ATS) phenomena have been investigated both theoretically and experimentally~\cite{PM2011,XG2016,WX2016,JL2018}. Meanwhile, it is also possible to control photon by photon, to realize an all optical routing~\cite{is2014} or microwave-photon detector~\cite{ki2016}.
Combined with the interference effect, it motivates us to study how to modulate the single photon scattering by the giant atom.}

In this paper, we tackle this issue in a {one-dimensional} waveguide with a two-level or three-level giant atom.
For the two-level atom setup, we find that the change in the size of the atom can control its energy shift due to the interference effect between the backward and forward photon in the waveguide.
{In the driven three-level giant-atom system, we demonstrate the controllable phase-modulated ATS physics. Our results show that, the size of the giant atom, which serves as a controller, can be used to tune the width of the transmission valleys (reflection range) in a periodical manner.}
The underlying physical principal is further revealed in the viewpoint of quantum open system based on the dressed state representation.

The rest of the paper is structured as follows:
In Sec.~\ref{twolevel}, we present the model and discuss the single photon scattering in a two-level {giant-atom} system.
In Sec.~\ref{threelevel}, we demonstrate the {Autler-Townes splitting} behavior for a driven three-level {giant-atom} with $\Lambda$-type transition. {In Sec.~\ref{dissipation}, we discuss the effects of the dissipation for both of two-level and three-level giant atom setups.} At last, we end up with a brief conclusion in Sec.~\ref{conclusion}.

\section{Two-level giant atom}
\label{twolevel}
\begin{figure}
\begin{centering}
\includegraphics[width=1\columnwidth]{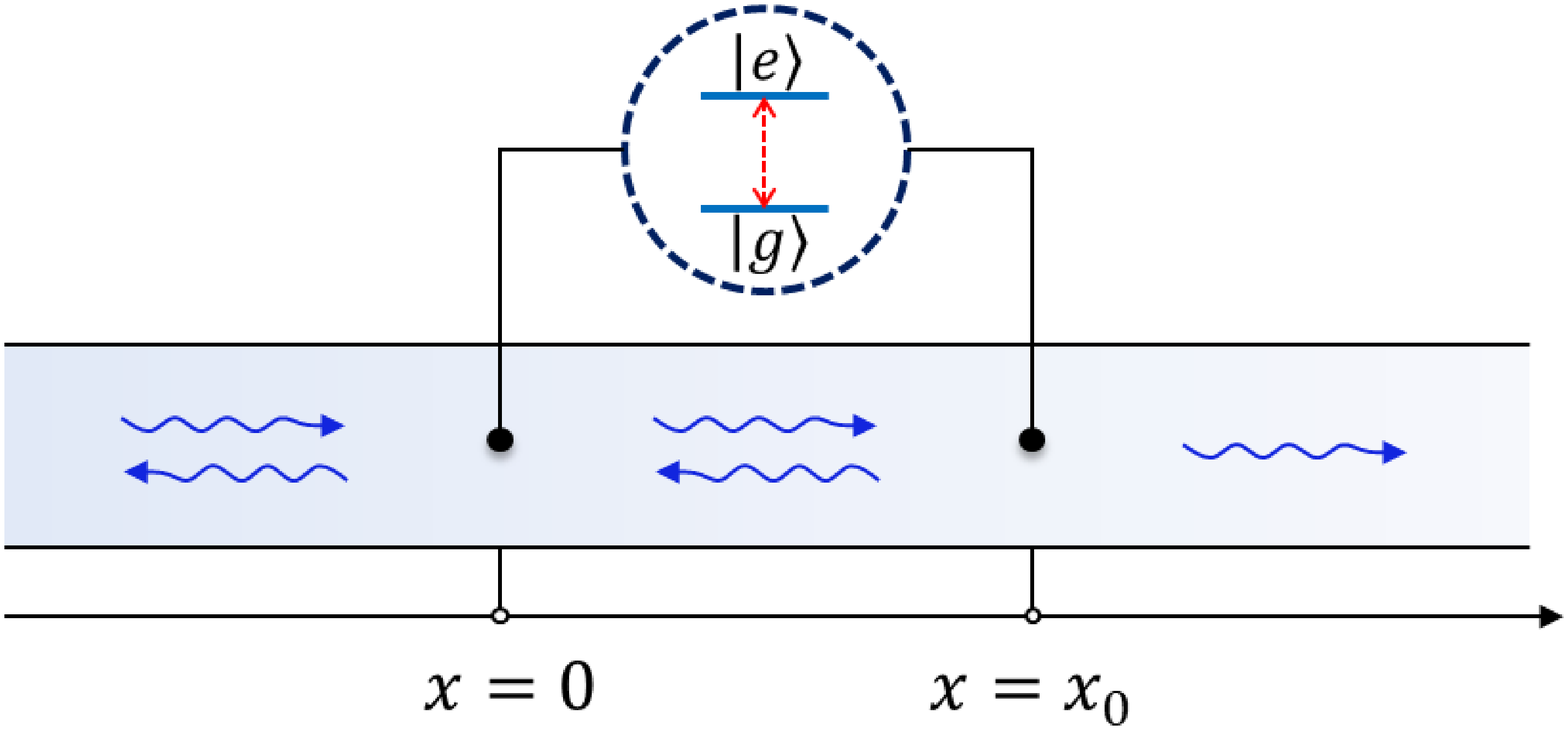}
\par\end{centering}
\caption{Schematic configuration for a linear waveguide coupled to a giant atom at the points $x=0$ and $x=x_0$.}
\label{f1}
\end{figure}

As schematically shown in Fig.~\ref{f1}, the system we consider is composed {of} a linear waveguide and a giant atom, which is actually a two-level system.
The giant atom is connected to the waveguide via two points with $x=0$ and $x=x_0${, respectively.}
The Hamiltonian $H_1$ of the system can be divided into three parts, i.e., $H_1=H_{s}+H_{\omega}+V_1$.
The first part $H_{s}$ is the free Hamiltonian of the giant atom (Hereafter, we set $\hbar=1$).
\begin{equation}
H_{s}=\omega_e\left|e\right\rangle \left\langle e\right|,
\end{equation}
where $\omega_e$ is the transition frequency between the ground state $\left|g\right\rangle $ and the excited state $\left|e\right\rangle $. As a reference, we have set the frequency of the ground state $\left|g\right\rangle $ as $\omega_g=0$.

The second part $H_{\omega}$ of the Hamiltonian $H_1$ represents the free Hamiltonian of the waveguide, and is expressed as
\begin{equation}
\label{eq:2}
H_{\omega}=\int dx\{ -iv_{g}C_{R}^{\dagger}\left(x\right)\frac{d}{dx}C_{R}\left(x\right)
+iv_{g}C_{L}^{\dagger}\left(x\right)\frac{d}{dx}C_{L}\left(x\right)\},
\end{equation}
where $v_{g}$ is the group velocity of photons traveling in the waveguide. $C_{R}^{\dagger}(x)[C_{L}^{\dagger}(x)]$ is the bosonic {creation} operator for the right-going (left-going) photon at position $x$.

For the third part $V_1$ of the Hamiltonian $H_1$, we describe the interaction between the waveguide and the giant atom.
Within the rotating wave approximation, the Hamiltonian $V_1$ can be expressed as
\begin{equation}
\begin{split}
V_1=&f\int dx\delta(x)\left[\sigma^{+}C_{R}\left(x\right)+\sigma^{+}C_{L}\left(x\right)+{\rm H.c.}\right]\\
&+f\int dx\delta\left(x-x_{0}\right)\left[\sigma^{+}C_{R}\left(x\right)+\sigma^{+}C_{L}\left(x\right)+{\rm H.c.}\right],
\end{split}
\end{equation}
where $f$ is the coupling strength between the waveguide and the two-level giant atom. $\sigma^+=(\sigma^-)^{\dagger}=|e\rangle\langle g|$ is the raising operator of the atom.
The Dirac-$\text{\ensuremath{\delta}}$ function in the Hamiltonian $V_1$ indicates that the giant atom has a length of $x_0$ and connects to the waveguide via its head and tail, that is, $x=0$ and $x=x_{0}$.

It is noted that, the total excitation of the atom and the photon in the waveguide is conserved.
In the following section, we will restrict {ourselves} in the single excitation subspace, to investigate how to control the single photon scattering state via adjusting the {frequency of the photon} and the size of the two-level giant atom.

In the single-excitation subspace, the eigenstate of the system can be written as
\begin{equation}
\begin{split}
\left|E\right\rangle =&\int dx\left[\phi_{R_1}\left(x\right)C_{R}^{\dagger}\left(x\right)+\phi_{L_1}\left(x\right)C_{L}^{\dagger
}\left(x\right)\right]\left|G\right\rangle\\
&+u_{e}\sigma^{+}\left|G\right\rangle,
\end{split}
\end{equation}
where $|G\rangle$ represents that the waveguide is in the vacuum {states while} the giant atom is in the ground state $|g\rangle$.
$\phi_{R_1}\left(x\right)$ and $\phi_{L_1}\left(x\right)$ are single-photon wave functions of the right-going and left-going modes in the waveguide{, respectively.}
$u_{e}$ is the excitation amplitude of the giant atom.
Solving the stationary Sch\"{o}dinger equation
{$H_{1}\left|E\right\rangle =E\left|E\right\rangle $}, the amplitudes equation can be obtained as
\begin{subequations}
\label{eq:5}
\begin{eqnarray}
-iv_{g}\frac{d}{dx}\phi_{R_1}\left(x\right)+fu_{e}M
&=&E\phi_{R_1}\left(x\right), \\
iv_{g}\frac{d}{dx}\phi_{L_1}\left(x\right)+fu_{e}M
&=&E\phi_{L_1}\left(x\right), \\
\omega_e u_{e}+fN&=&Eu_{e}.
\end{eqnarray}
\end{subequations}
where $M=\delta(x)+\delta(x-x_{0})$ and $N=\phi_{R_1}(0)+\phi_{R_1}(x_{0})+\phi_{L_1}(0)
+\phi_{L_1}(x_{0})$.

Next, we consider the scattering behavior when a single photon with wave vector $k$ {is incident} from the left side of the waveguide.
In this case, the wave functions of $\phi_{R_1}\left(x\right)$ and $\phi_{L_1}\left(x\right)$ can be expressed as
\begin{equation}
\label{eq:6}
\begin{split}
\phi_{R_1}\left(x\right)=&e^{ikx}\{ \theta\left(-x\right)+A_1\left[\theta\left(x\right)-\theta\left(x-x_{0}\right)\right]\\&
+t_1\theta\left(x-x_{0}\right)\},
\end{split}
\end{equation}
\begin{equation}
\label{eq:7}
\phi_{L_1}\left(x\right)=e^{-ikx}\left\{ r_1\theta\left(-x\right)+B_1\left[\theta\left(x\right)-\theta\left(x-x_{0}\right)\right]\right\},
\end{equation}
with
\begin{equation}
\label{eq:8}
\theta\left(x\right)=\begin{cases}
1 & x>0\\
\frac{1}{2} & x=0\\
0 & x<0
\end{cases}.
\end{equation}
We explain the expression of the above wave functions physically as follows.
When the right-going photon incident from the region $x<0$ reaches the first connection point $x=0$ between the giant atom and the waveguide, it can be transmitted or reflected, with the amplitudes of $A_1$ and {$r_1$,} respectively.
The {photon} transmitted at the first {connection} point at $x=0$ will travel freely in the waveguide until it reaches the second {connection} point  at $x=x_0$, it will be then reflected or transmitted secondly, with the amplitudes $B_1$ and $t_1$, respectively.

Now, we substitute Eq.~(\ref{eq:6}-\ref{eq:8}) into the amplitude Eq.~(\ref{eq:5}), it yields the dispersion relation $E=v_{g}k$, and
\begin{equation}
t_{1}=\frac{i\Delta v_{g}-2if^{2}\sin \left(kx_{0}\right)}{i\Delta v_{g}-2f^{2}\left(1+e^{ikx_{0}}\right)}.
\end{equation}
where $\Delta=E-\omega_e$ is the detuning between the atom and the propagating photon in the waveguide.

Furthermore, the transmission rate $T_1=\left|t_1\right|^{2}$ can be obtained as
\begin{equation}
\label{eq:9}
T_1=\frac{\left(\Delta v_{g}-2f^{2}\sin \left(kx_{0}\right)\right)^{2}}{\left(\Delta v_{g}-2f^{2}\sin \left(kx_{0}\right)\right)^{2}+4f^{4}\left(1+\cos \left(kx_{0}\right)\right)^{2}},
\end{equation}

\begin{figure}
\begin{center}
\includegraphics[width=1\linewidth]{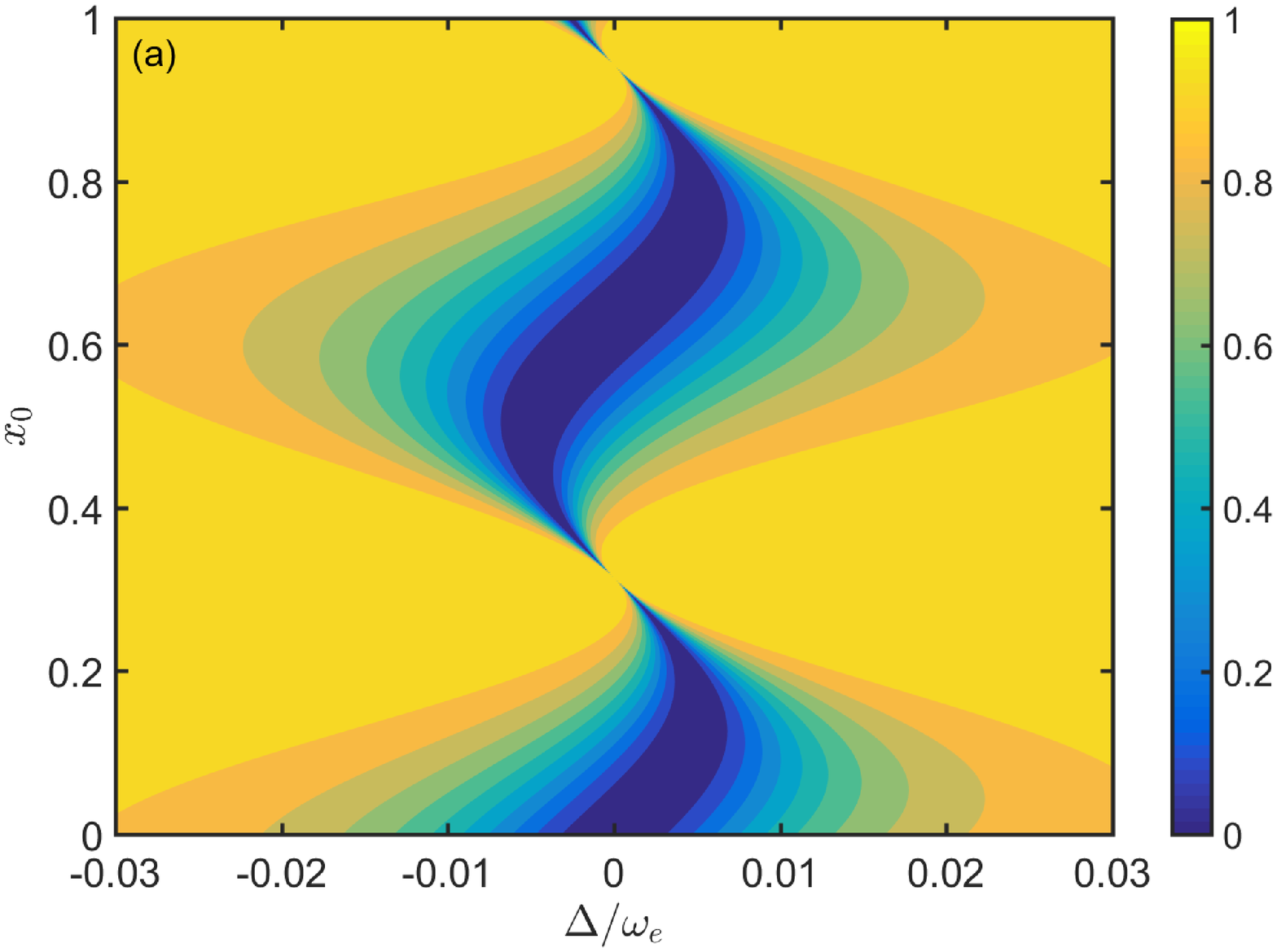}
\includegraphics[width=1\linewidth]{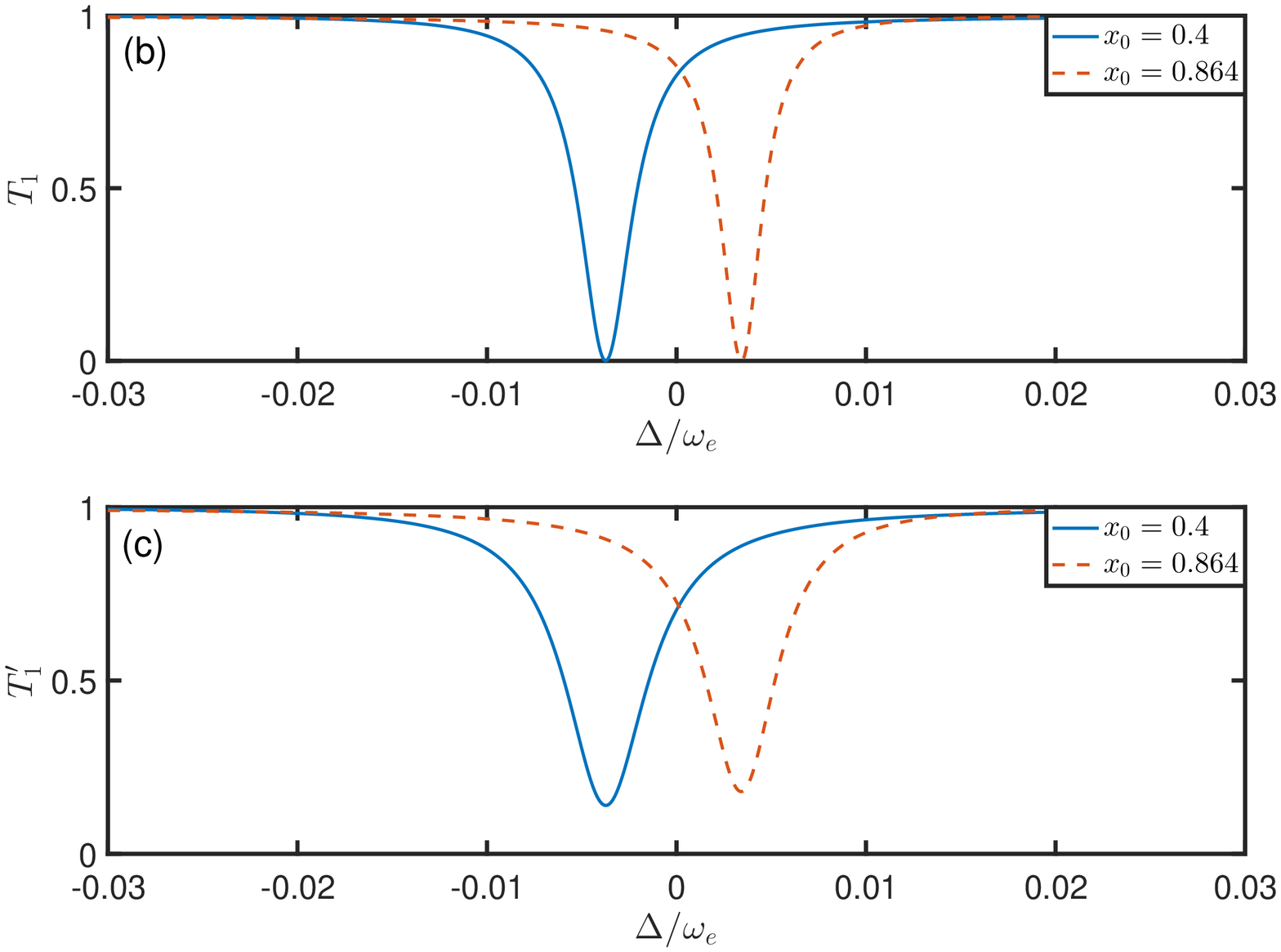}
\end{center}
\caption{(a) The transmission rate $T_1$ as functions of $x_0$ and $\Delta$. (b) The transmission rate $T_1$ as a function of $\Delta$. (c) The transmission rate $T_{1}^{\prime}$ as a function of $\Delta$. The parameters are set as $\omega_e=3$\,GHz, $v_g=3\times10^8$\,m/s, $f/\sqrt{v_{g}\omega_{e}}=0.05$. The other parameters is $\gamma_{e}=10^{-3}\omega_{e}$ for (c).}
\label{f2}
\end{figure}

In the small atom scenario ($x_0=0$) which is studied in Ref.~\cite{jt2005}, it is obvious that the incident photon will be completely reflected ($T_1=0$) when it is resonant with the atom, that is, $\Delta=0$.
However, for the giant atom in our setup, the incident photon will propagate back and forth in the spatial regime covered by the giant atom, leading to an interference effect. As a result, as shown in Fig.~\ref{f2}(a), the transmission rate can be controlled by adjusting the photon-atom detuning $\Delta$ and the size of the giant atom $x_0$.

Furthermore, considering the atomic spontaneous emission, where we replace $\omega_e$ by $\omega_e-i\gamma_e$,  the transmission amplitudes $t_{1}^{\prime}$ becomes
\begin{equation}
t_{1}^{\prime}=\frac{i\left(\Delta+i\gamma_{e}\right)v_{g}-2if^{2}\sin \left(kx_{0}\right)}{i\left(\Delta+i\gamma_{e}\right)v_{g}-2f^{2}\left(1+e^{ikx_{0}}\right)},
\end{equation}
where $\gamma_{e}$ is the spontaneous emission of the giant atom. As shown in Fig.~\ref{f2}(b) and (c), we plot the transmission rates $T_1$ and $T_{1}^{\prime}=\left|t_{1}^{\prime}\right|^{2}$ as functions of the detuning $\Delta$, respectively, to compare the results with and without spontaneous emission. It shows that the spontaneous emission breaks the complete transmission and makes the transmission valleys become wider and shallower. However, the locations of the valleys are not changed. It implies that the environment nearly doesn't  change the energy structure of the system.

\begin{figure}
\begin{center}
\includegraphics[width=1\linewidth]{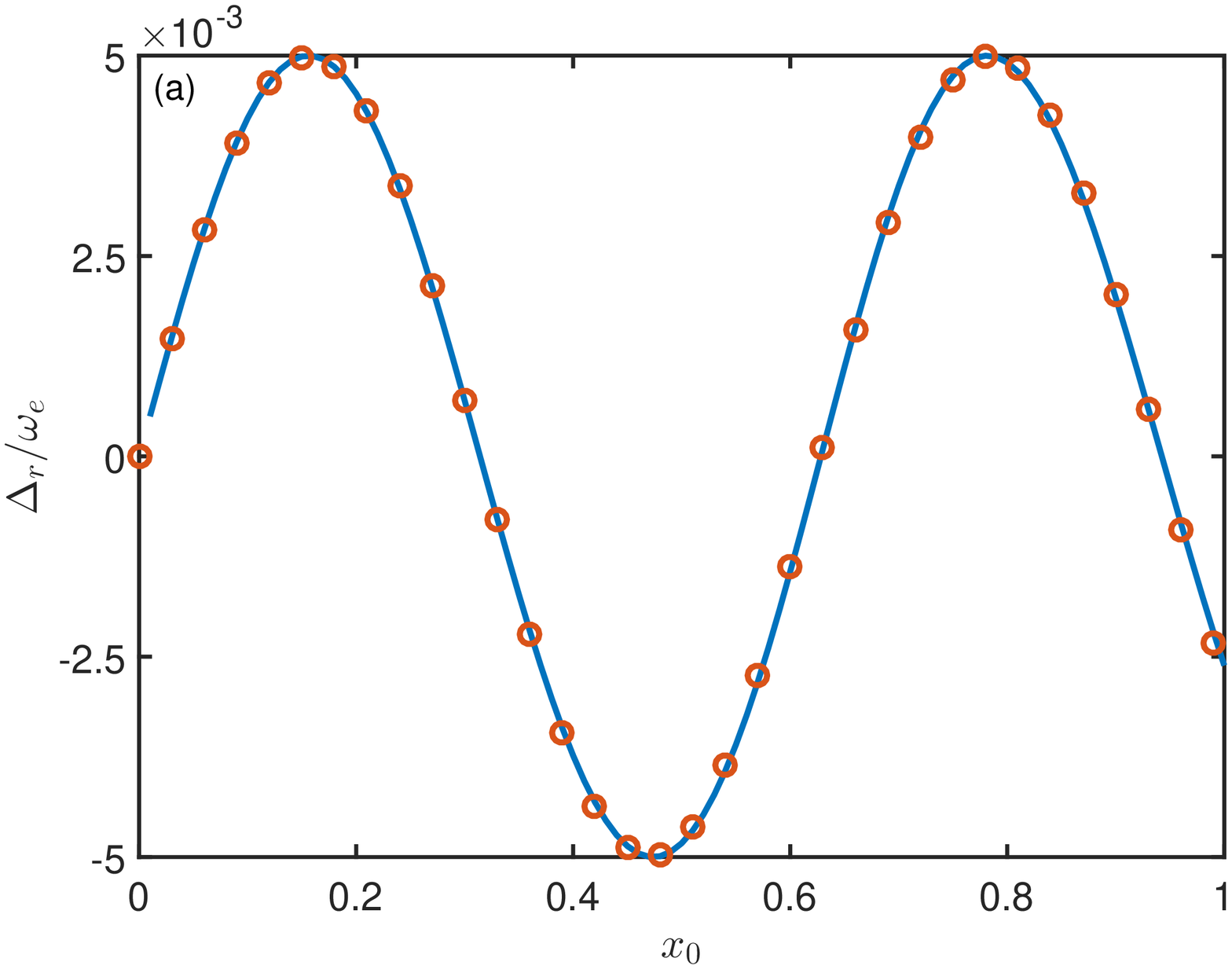}
\includegraphics[width=1\linewidth]{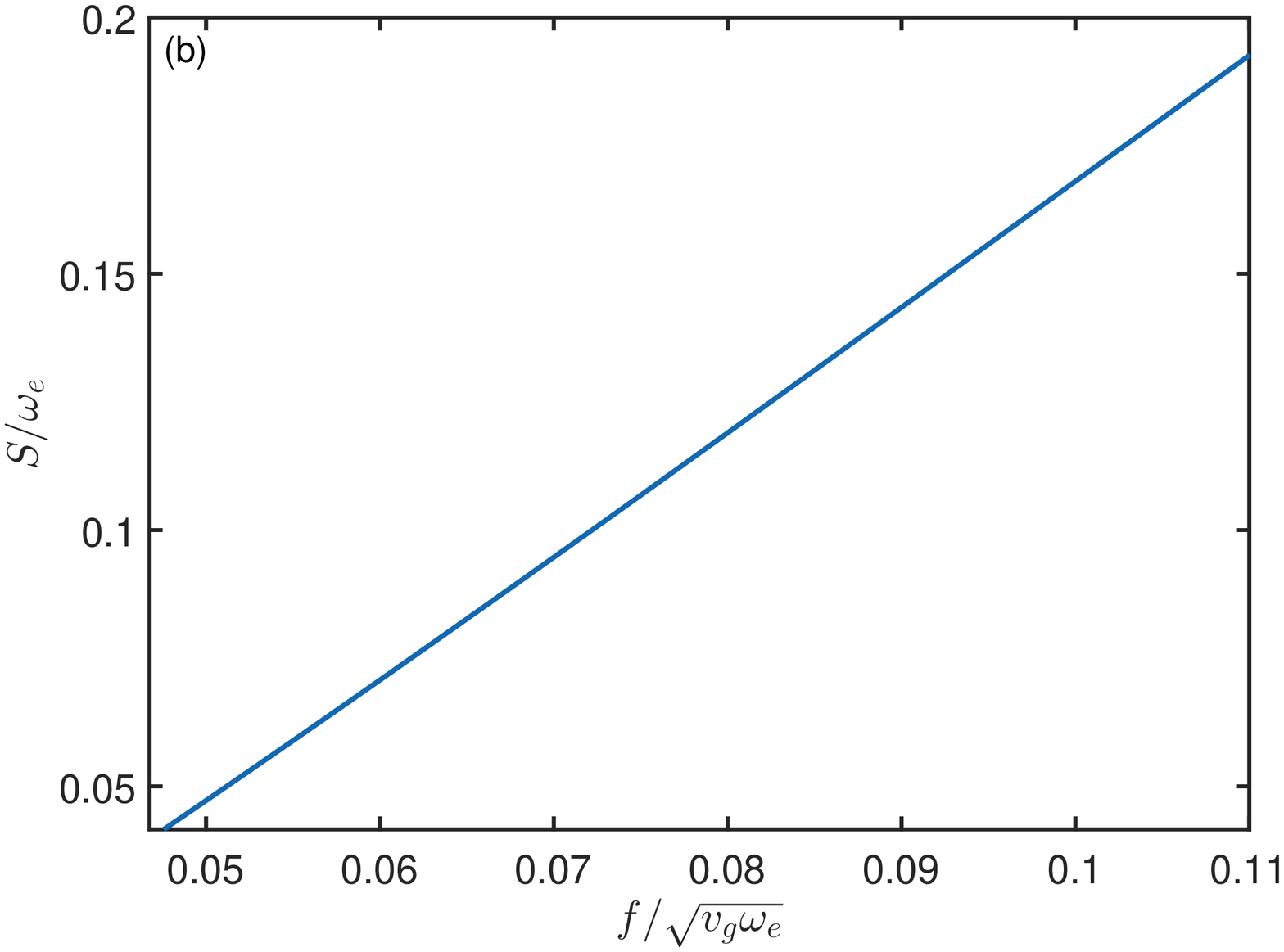}
\end{center}
\caption{(a) The detuning $\Delta_r$ as {a} function of $x_0$ for $\omega_e=3$\,GHz, $v_g=3\times10^8$\,m/s, $f/\sqrt{v_{g}\omega_{e}}=0.05$.
The solid line is given by the solution of the transcendental equation in Eq.~(\ref{tre}) and the empty circles are the results from the numerical fitting.
(b) The fitting parameters $S$ in Eq.~(\ref{fit}) as a function of the atom-waveguide coupling strength $f$.}
\label{f3}
\end{figure}

More interestingly, in the giant-atom situation ($x_0\neq0$), the detuning for complete reflection is determined by the transcendental equation
\begin{equation}
\Delta_r=\frac{2f^{2}\sin \left(k_rx_{0}\right)}{v_{g}},
\label{tre}
\end{equation}
where $\Delta_r$ is the detuning between the atom and the propagating photon when the incident photon is completely reflected, and $k_r$ is the corresponding photonic wave vector, $\Delta_r$ is dependent on $k_r$ via $\Delta_r=v_gk_r-\omega_e$. It is well known that the atomic frequency is determined by the location of the complete reflection of the incident photon~\cite{jt2005}. Therefore, in Fig.~\ref{f2}(b), we can observe an atomic frequency shift which originates from the interference effect as discussed above, compared {with} the small atom system. Since $\left|\sin k_{r}x_{0}\right|\leq1$, Eq.~(\ref{tre}) implies that only when the frequency of the incident photon satisfies $|\Delta_r|\leq2f^2/v_g$, that is, $\omega_e+2f^2/v_g\geq v_gk_r\geq\omega_e-2f^2/v_g$, it is possible to be completely reflected. Within this regime, we plot the dependence of $\Delta_r$ on the size of giant atom $x_0$ in Fig.~\ref{f3}(a) (solid line), which shows a sinusoidal shape. The exact sinusoidal shape by the numerical fitting (empty circles) is also shown in the figure and the fitting function is obtained by
\begin{equation}
\Delta_r\approx S\sin(\omega_{e}x_{0}/v_g),
\label{fit}
\end{equation}
where the fitting parameters $S\approx 2f^2/v_g$ is plotted as a function of atom-waveguide coupling strength $f$ in Fig.~\ref{f3}(b). We recall that, when two identical small atoms are coupled to the waveguide at the position $x=0$ and $x_0$, the photon will also propagate back and forth between the two atoms. However, as shown in Appendix, we can not observe the shift, that is $T=0$ occurs at $\Delta=0$. The reason is that, small atoms do not interfere with themselves. However, in giant atoms, the interference within an atom promotes the apparent frequency shift in its energy levels.

In experiments, the giant atom setup can be realized by the superconducting circuits, and the transition frequency is in the order of GHz~\cite{bk2020,AM2020}, while the intrinsic decoherence time of the superconducting qubit is achieved by $20{\rm \mu s}$ ~\cite{FY2017} or even longer ~\cite{AJ2017}. The natural line width of the giant atom is then $\gamma=0.05$\,MHz . In Fig.~\ref{f3}(a), the energy shift is with the regime of $|\Delta_r|=5\times10^{-3}\omega_{e}=15$\,MHz, for some $x_0$ so that $\Delta_{r}\gg\gamma$, and $\Delta_{r}$ is about $300$ times larger than $\gamma$. Therefore, the energy shift of the giant atom can be larger than its natural line width and observed experimentally by increasing the atom-waveguide coupling strength.

\section{THREE-LEVEL giant atom}
\label{threelevel}
\begin{figure}
\begin{centering}
\includegraphics[width=1\columnwidth]{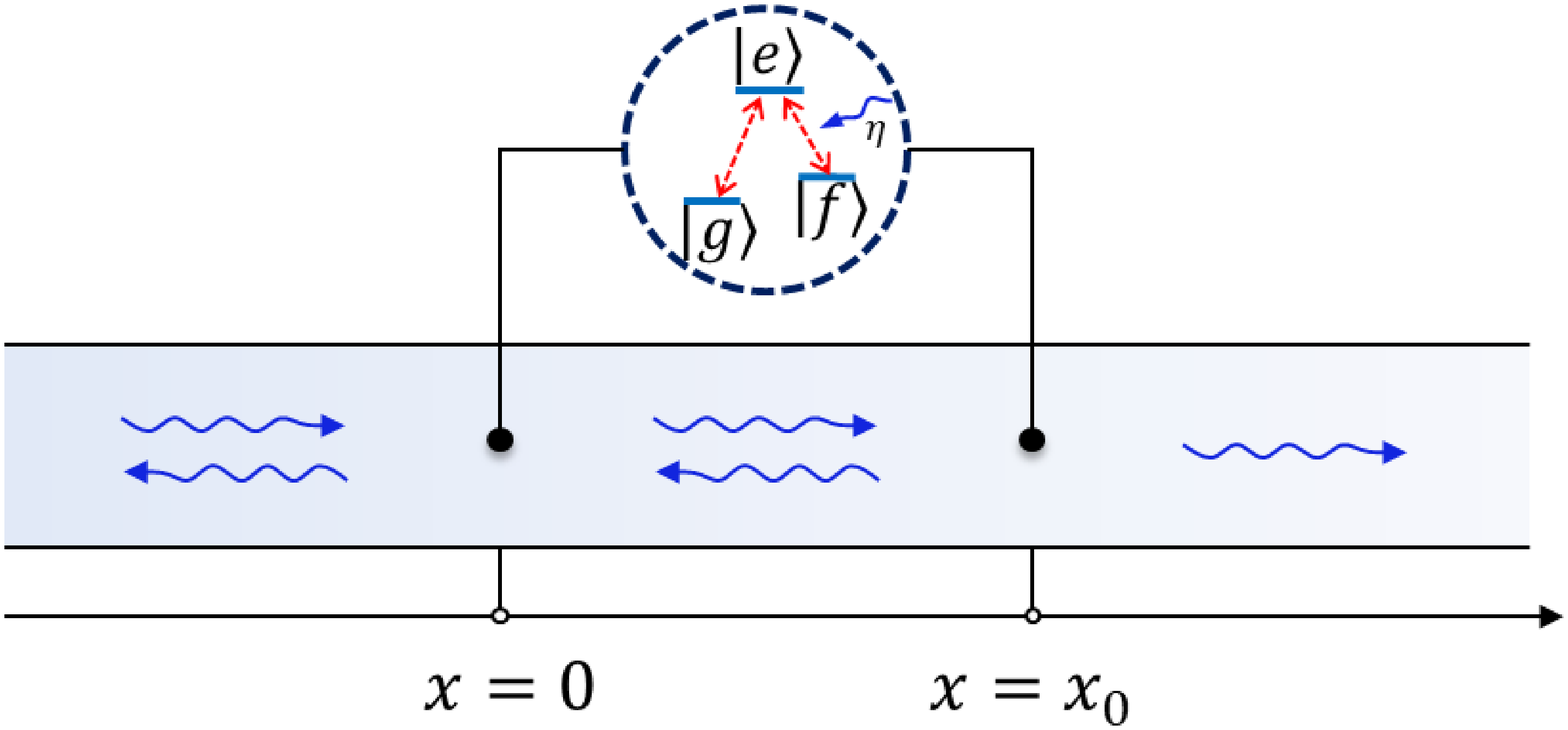}
\par\end{centering}
\caption{Schematic configuration for a linear waveguide coupled to a $\Lambda$-type three-level giant atom at the points $x=0$ and $x=x_0$.}
\label{f4}
\end{figure}

{In this section, let us consider the single-photon scattering on a three-level giant atom driven in $\Lambda$ type.
As schematically shown in Fig.~\ref{f4}, the atom is characterized by the ground state $\left|g\right\rangle$, excited state $\left|e\right\rangle$ and metastable state $\left|f\right\rangle$.
Here, there are two types of physical processes in this system.
One is similar to that in two-level giant atom system, that is, the interplay between the multiple backward and forward photons of the transmission and reflection in the waveguide;
the other is peculiar for such a three-level atomic system, i.e., that the strong driving field couples the states $|e\rangle$ and $|f\rangle$ and a weak field (the incident photon in the waveguide) probes the states $|e\rangle$ and $|g\rangle$ will induce the ATS phenomenon.
The incorporation between these two processes may provide a possibility of the extra and interesting single-photon scattering spectrum for the giant-atom setup.}

The Hamiltonian $H_2$ of the current system can be divided into three parts, i.e., $H_2=H_{w}+H_{a}+V_2$.
Here, $H_w$ is the free Hamiltonian of the waveguide which is given in Eq.~(\ref{eq:2}). $H_{a}$ is the Hamiltonian of the giant atom
\begin{equation}
\begin{split}
H_{a}=&\omega_{e}\left|e\right\rangle \left\langle e\right|+\omega_{f}\left|f\right\rangle \left\langle f\right|\\
&+\eta(\left|e\right\rangle \left\langle f\right|e^{-i\omega_{d}t}+\left|f\right\rangle \left\langle e\right|e^{i\omega_{d}t}),
\end{split}
\end{equation}
where $\omega_{e}$ and $\omega_{f}$  ($\omega_{f}<\omega_{e}$) are the frequencies of the state $|e\rangle$ and $|f\rangle$, respectively.
As a reference, we have set $\omega_g=0$.
$\eta$ and $\omega_d$ are{, respectively,} the strength and frequency of the classical field, which drives $|f\rangle\leftrightarrow|e\rangle$ transition.  In the rotating frame, the time independent Hamiltonian becomes
\begin{equation}
\tilde{H}_{a}=\omega_{e}\left|e\right\rangle \left\langle e\right|+\delta\left|f\right\rangle \left\langle f\right|
+\eta(\left|e\right\rangle \left\langle f\right|+\left|f\right\rangle \left\langle e\right|),
\label{rotating}
\end{equation}
where $\delta=\omega_{f}+\omega_d$.

The third part $V_2$ of the Hamiltonian $H_2$  describes the interaction between the waveguide and the giant atom. Within the rotating wave approximation, it can be expressed as
\begin{equation}
\begin{split}
V_{2}&=f\int dx\delta(x)\left\{ \left[C_{R}^{\dagger}(x)+C_{L}^{\dagger}(x)\right]\left|g\right\rangle \left\langle e\right|+{\rm H.c.}\right\} \\
&+f\int dx\delta(x-x_{0})\left\{ \left[C_{R}^{\dagger}(x)+C_{L}^{\dagger}(x)\right]\left|g\right\rangle \left\langle e\right|+{\rm H.c.}\right\},
\end{split}
\end{equation}
where $f$ is the coupling strength between the waveguide and giant atom.

In the single-excitation subspace, the eigenstate of the system can be expressed as
\begin{equation}
\begin{split}
\left|E\right\rangle =&\int dx\left[\phi_{R_2}(x)C_{R}^{\dagger}+\phi_{L_2}(x)C_{L}^{\dagger}\right]\left|\emptyset,g\right\rangle\\ &+\lambda_{e}\left|\emptyset,e\right\rangle +\lambda_{f}\left|\emptyset,f\right\rangle,
\end{split}
\end{equation}
where $|\emptyset,m\rangle (m=e,g,f)$ represents that the waveguide is in the vacuum state {while} the giant atom is in the state $|m\rangle$. $\phi_{R_2}\left(x\right)$ and $\phi_{L_2}\left(x\right)$ are single-photon wave functions of the right-going and left-going modes in the waveguide{, respectively.} $\lambda_{e}$ and $\lambda_{f}$ are the excitation amplitudes of the giant atom in the excited state $\left|e\right\rangle$ and metastable state $\left|f\right\rangle$, respectively.
Similar to the discussion in two-level atom setup, for a {right-}going incident photon with wave vector $k$, the photon amplitude can be expressed as
\begin{equation}
\label{eq:15}
\begin{split}
\phi_{R_2}\left(x\right)=&e^{ikx}\{ \theta\left(-x\right)+A_2\left[\theta\left(x\right)-\theta\left(x-x_{0}\right)\right]\\&
+t_2\theta\left(x-x_{0}\right)\},
\end{split}
\end{equation}
\begin{equation}
\label{eq:16}
\phi_{L_2}\left(x\right)=e^{-ikx}\left\{ r_2\theta\left(-x\right)+B_2\left[\theta\left(x\right)
-\theta\left(x-x_{0}\right)\right]\right\},
\end{equation}
where $t_2$ and $r_2$ are the transmission and reflection amplitudes while $A_2$ ($B_2$) are the amplitudes for finding a right (left)-going photon inside the regime of the giant atom. Then, the Sch\"{o}dinger equation $\tilde{H}|E\rangle=E|E\rangle$ (where $\tilde{H}=H_w+\tilde{H}_a+V_2$) yields

\begin{equation}
t_{2}=\frac{(E-\delta)\left[i\left(E-\omega_{e}\right)v_{g}-2if^{2}
\sin\left(kx_{0}\right)\right]-iv_{g}\eta^{2}}{(E
-\delta)\left[i\left(E-\omega_{e}\right)v_{g}-2f^{2}\left(1+e^{ikx_{0}}\right)\right]-iv_{g}\eta^{2}},
\label{tt}
\end{equation}
\begin{equation}
r_{2}=\frac{2f^{2}(E-\delta)\left(1+\cos\left(kx_{0}\right)
\right)e^{ikx_{0}}}{(E-\delta)\left[i\left(E
-\omega_{e}\right)v_{g}-2f^{2}\left(1+e^{ikx_{0}}\right)\right]-iv_{g}\eta^{2}}\label{rr}.
\end{equation}
Then the transmission rate $T_2=|t_2|^2$ and reflection rate $R_2=|r_2|^2$ satisfy $T_2+R_2=1$ due to the neglect of the natural relaxations in the atom.

In Fig.~\ref{f7}, we plot the transmission rate $T_2$ as functions of the detuning $\Delta=E-\omega_e$ {between the incident photon and atomic $|g\rangle\leftrightarrow|e\rangle$ transition and the size of the giant atom $x_0$.}
Here, we illustrate the result for {resonantly driving the atom by the field $\eta$} in Fig.~\ref{f7}(a) and (d), and {non-resonantly} in Fig.~\ref{f7}(b),(c) and (e).
All of the results are characterized by {two narrow transmission valleys ($T_2\simeq 0$) and a relatively wide transmission windows ($T_2\simeq 1$) between them around the two-photon resonance.

{First, we focus on the transmission window {($T_2\simeq 1,R_2\simeq 0$).} It can be observed in Eqs.~(\ref{rr}) that {$R_2$=0} when $E-\delta=E-\omega_f-\omega_d=0$, which implies the two-photon resonance condition $\omega_e-E=\omega_e-\omega_f-\omega_d$. This transmission window is caused by the usual ATS mechanism, which is true for both
of the small and giant atom setup. Moreover, we can also observe that {$R_2=0$} when $1+\cos{kx_0}=0$, that is $kx_0=(2m+1)\pi$ where $m$ is an integer, and {the dependence of the position} is peculiar for the giant atom setup. Note that $kx_0$ is the accumulated phase as the travelling photon moves from one coupling point to the other. {Therefore, such a ATS transmission is also related to the position-dependent phase introduced by the interference effect from the back and forth photons inside the regime covered by the giant atom.}}
\begin{figure*}
\begin{centering}
\subfigure{
\includegraphics[width=1\columnwidth]{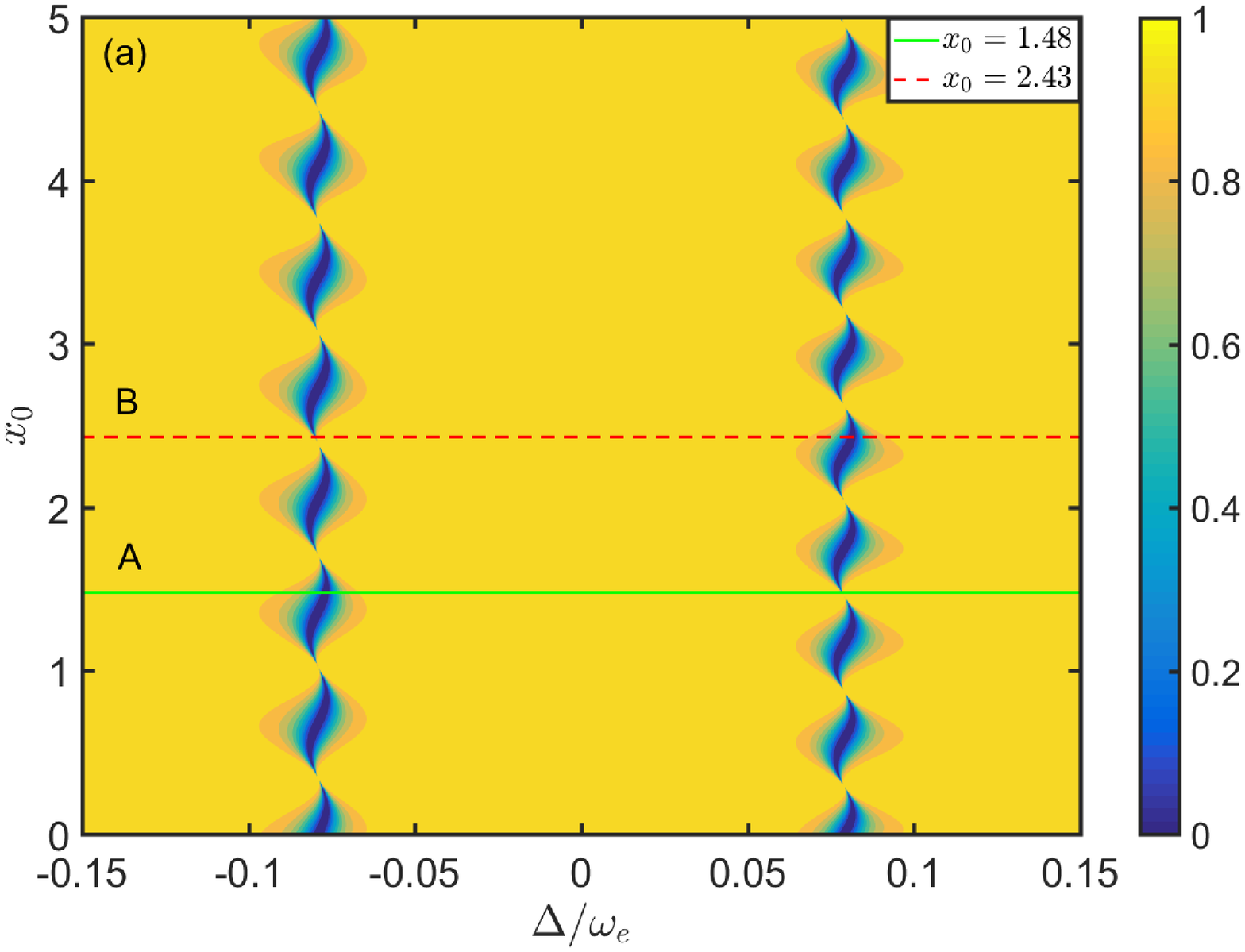}}
\subfigure{
\includegraphics[width=1\columnwidth]{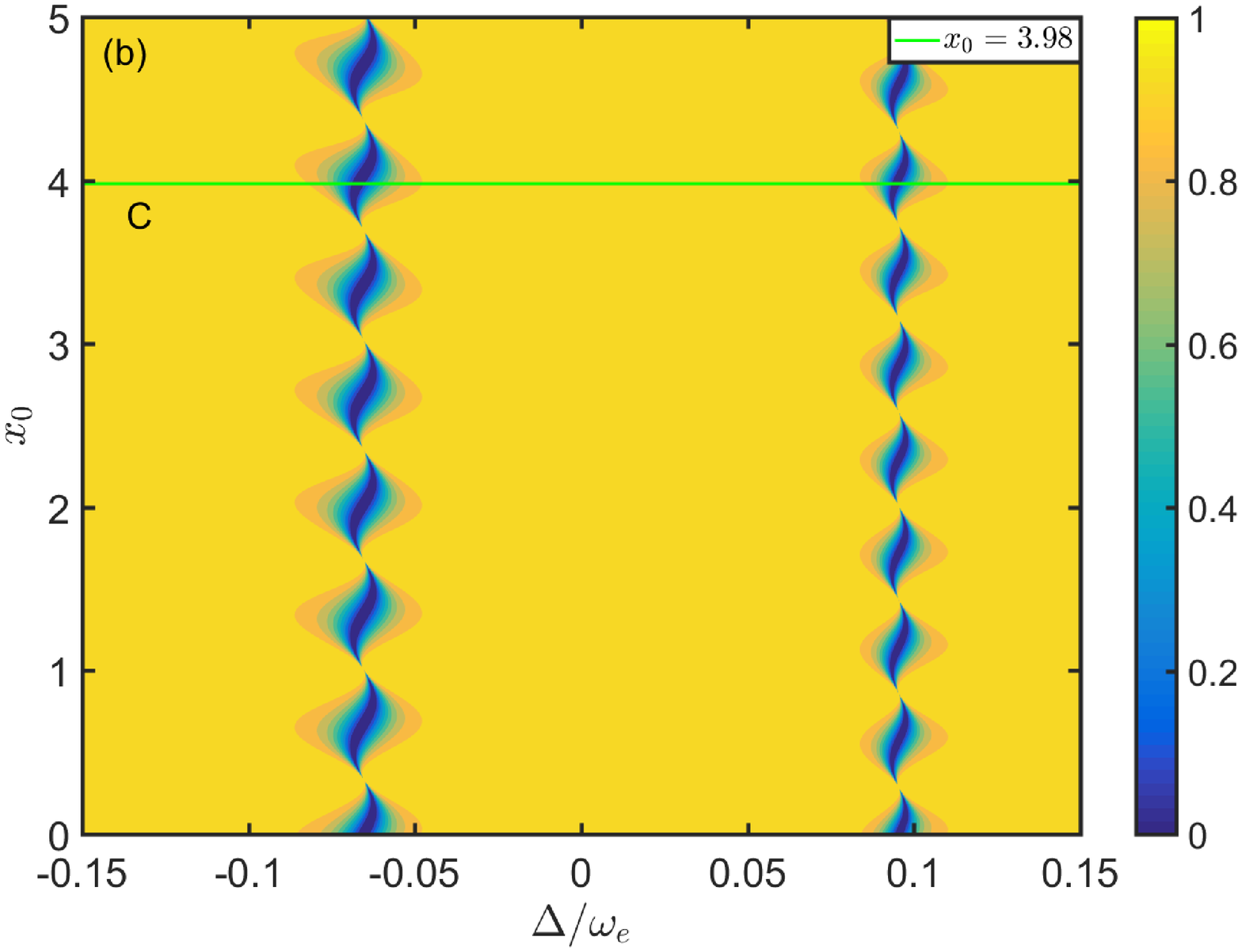}}
\subfigure{
\includegraphics[width=1\columnwidth]{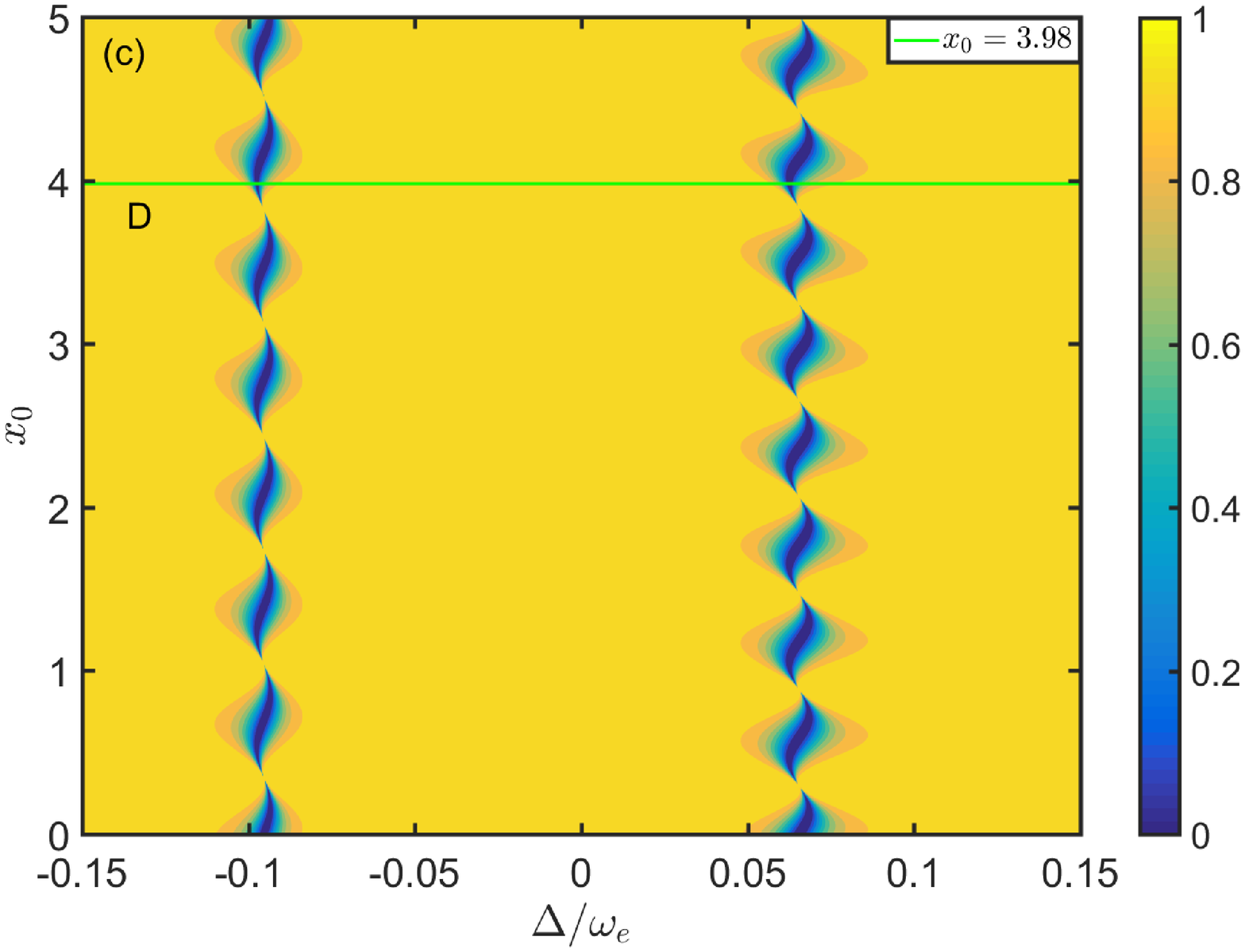}}
\subfigure{
\includegraphics[width=1\columnwidth]{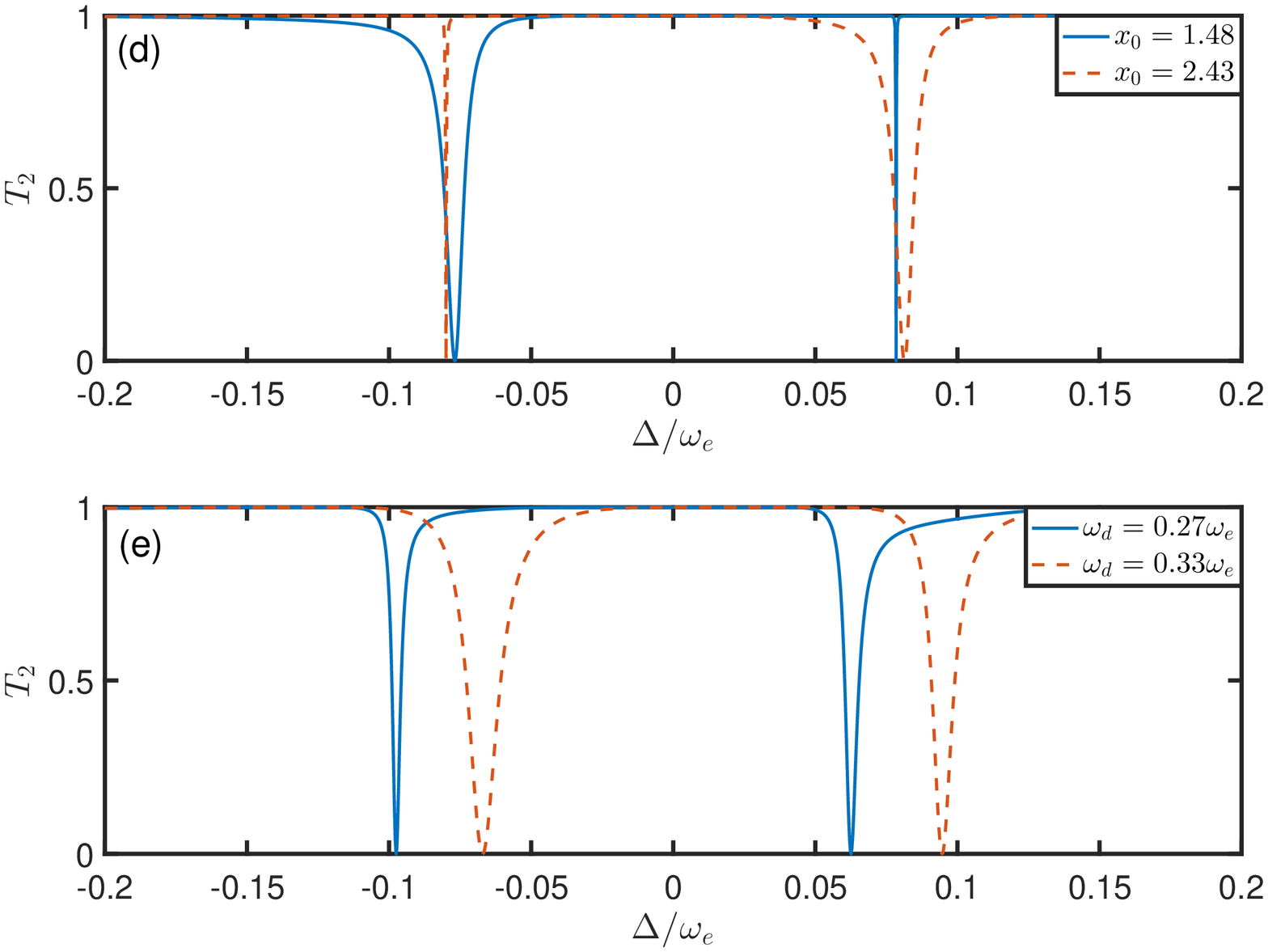}}
\end{centering}
\caption{The transmission rate $T_2$ as functions of the detuning $\Delta=E-\omega_e$ and the size of the giant atom $x_0$ for resonant driving.  The parameters are set as $\omega_e=3$\,GHz, $v_g=3\times10^8$\,m/s, $\omega_f=0.7\omega_e$, $f/\sqrt{v_{g}\omega_{e}}=0.05$, $\eta=5f=0.08\omega_e$. The other parameters are $\omega_d=0.3\omega_e$, $\omega_d=0.33\omega_e$, $\omega_d=0.27\omega_e$ for (a), (b) and (c){, respectively.} (d) The transmission rate $T_2$  for the lines labelled by ``A" (solid) and ``B" (dashed) in (a). (e) The transmission rate $T_2$ for the line labelled by ``C" (solid) and ``D" (dashed) in (b) and (c), respectively.}
\label{f7}
\end{figure*}

\begin{table*}
  \centering
  \caption{The values of $x_0$, $\Delta_2$ and corresponding values of $|G_{\pm,{k}}|$ for the two valleys in Fig.~6 (d) and (e).}
\begin{tabular}{|c|c|c|c|c|c|}
\hline
  \hline
  Line Label&curve &$\Delta_2/\omega_e$ &$x_0$ & $|G_{-,\omega_{-}/v_g}|/f$&$|G_{+,\omega_{+}/v_g}|/f$ \\
  \hline
  A &solid in Fig.~~\ref{f7}(d) &$0$ & $1.48$& $1.2069$ & $0.1783$ \\
    \hline
  B &dashed in Fig.~~\ref{f7}(d)& $0$ & $2.43$ & $0.2574$ & $1.1892$ \\
    \hline
  C &dashed in Fig.~~\ref{f7}(e)& $-0.03$ & $3.98$ & $1.4813$ & $1.2469$\\
  \hline
  D &solid in Fig.~~\ref{f7}(e)& $0.03$ & $3.98$ & $0.7944$ & $1.0316$\\
  \hline
\end{tabular}
\label{parameters}
\end{table*}

Next, we discuss the two valleys, which represent the complete reflection {$(T_2\simeq0,R_2\simeq1)$.}
Based on Eq.~(\ref{tt}), the complete reflection occurs when
\begin{eqnarray}
E_{\pm}&=&\omega_{e}+f^{2}\sin(kx_{0})/v_g-\frac{\Delta_2}{2}\nonumber \\ &&\pm\frac{\sqrt{\left(\Delta_2+2f^{2}\sin\left(kx_{0}\right)/v_g\right)^{2}+4\eta^{2}}}{2},
\label{valley}
\end{eqnarray}
where $\Delta_2=\omega_{e}-\omega_{f}-\omega_{d}$ is the detuning between the driving field and the atomic $|f\rangle\leftrightarrow|e\rangle$ transition.
This fact can be explained intuitively in the dressed state presentation.
The {eigenfrequencies} of the driving Hamiltonian $\tilde{H}_a$ in Eqs.~(\ref{rotating}) are
\begin{equation}
\omega_\pm=\omega_e-\frac{\Delta_2}{2}\pm\frac{\sqrt{\Delta_2^2+4\eta^2}}{2}.
\label{eigen}
\end{equation}
and the corresponding states are
\begin{eqnarray}
|\psi_+\rangle&=&\cos{\frac{\theta}{2}}|e\rangle+\sin{\frac{\theta}{2}}|f\rangle,\\
|\psi_-\rangle&=&-\sin{\frac{\theta}{2}}|e\rangle+\cos{\frac{\theta}{2}}|f\rangle.
\end{eqnarray}
For the {resonantly} driving ($\Delta_2=0$), we will have $\theta=\pi/2$. For the case of non-resonantly driving, we will have $\theta={\rm atan}(2\eta/\Delta_2)$ for $\Delta_2>0$ and $\theta=\pi+{\rm atan }(2\eta/\Delta_2)$ for $\Delta_2<0$.

Comparing Eqs.~(\ref{eigen}) with (\ref{valley}), we find that the incident photon will be completely reflected when it is ``nearly" resonant with $\omega_\pm$.
This fact is verified by the results shown in Fig.~\ref{f7}.
Here, we use the phrase ``nearly" to imply that $E_\pm$ is not exactly equal to $\omega_\pm$, but is slightly modulated by $x_0$.
As a result, the photon transmission shows a {periodic phase modulation in terms of $x_0$,} which is clearly demonstrated in Figs.~\ref{f7} (a), (b) and (c) for both of resonantly and non-resonantly driving situations.

It is also shown in Fig.~\ref{f7} that the two valleys discussed above possess different widths. For example, see the horizontal lines labelled by ``A", ``B", ``C" and ``D" in Figs.~\ref{f7} (a), (b) and (c). This can be explained by the different {effective decay rates} of the {eigenstates} with frequency $\omega_\pm$ in the viewpoint of quantum open system by regarding the waveguide as the effective environment.
To this end, we rewrite the interaction Hamiltonian $V_2$ in the momentum space by performing the Fourier transformation (in terms of $|\psi_\pm\rangle$) as
\begin{equation}
V_2=\sum_k\{[C_L^\dagger(k)+C_R^\dagger(k)]|g\rangle[G_{+,k}\langle\psi_+|+G_{-,k}\langle\psi_-|]
+{\rm H.c.}\}
\end{equation}
where
\begin{equation}
G_{+,k}=f\cos{\frac{\theta}{2}}(1+e^{ikx_0}),G_{-,k}=-f\sin{\frac{\theta}{2}}(1+e^{ikx_0}),
\end{equation}
characterize the coupling strength between the $k$th mode in the waveguide and the giant atom. We note that, $|G_{+,k}|$ and $|G_{-,k}|$ actually reflect the width of the two valleys when the value of $k$ is taken to satisfy $v_g k=\omega_\pm$.

In Table I, we list the values of $|G_{+,k}|$ and $|G_{-,k}|$  for the horizontal lines in Fig.~\ref{f7}(a) (b) and (c), along which the transmission rates are plotted as functions of the detuning $\Delta$ in Fig.~\ref{f7} (d) and (e). It shows a good agreement for the valley width between the table and the curves. For the resonant driving, which valley is wider depends
on the size of the giant atom $x_0$, as shown in Fig.~\ref{f7} (a) and (d). However, for the non-resonant driving, as shown in Fig~\ref{f7} (b), (c) and (e), the left valley is nearly always narrower than the right one when $\Delta_2>0$ and wider when $\Delta_2<0$. In this sense, the modification to the photon transmission by the atomic size is more sensitive for the resonant driving setup.

\begin{figure}
\includegraphics[width=1\columnwidth]{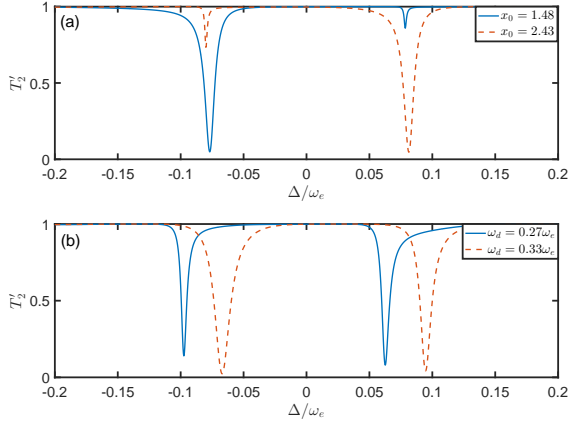}
\caption{The transmission rate $T_{2}^{\prime}=|t_{2}^{\prime}|^2$ as a function of $\Delta$ within the atomic spontaneous emission for resonant driving (a) and non-resonant driving (b). The parameters are set as $\omega_e=3$\,GHz, $v_g=3\times10^8$\,m/s, $\omega_f=0.7\omega_e$, $f/\sqrt{v_{g}\omega_{e}}=0.05$, $\eta=5f=0.08\omega_e$, $\gamma_e=\gamma_f=10^{-3}\omega_{e}$. The frequency of the driving field is set as $\omega_d=0.3\omega_e$ in (a). }
\label{f8}
\end{figure}

\section{Effects of dissipation}
\label{dissipation}
When the natural relaxation of atoms is taken into consideration, the transmission amplitude is expressed as
\begin{equation}
t_{2}^\prime=\frac{\Delta_f\left[i\left(E-\omega_{e}+i\gamma_{e}\right)v_{g}-2if^{2}
\sin\left(kx_{0}\right)\right]-iv_{g}\eta^{2}}{\Delta_f\left[i\left(E-\omega_{e}+i\gamma_{e}\right)v_{g}-2f^{2}\left(1+e^{ikx_{0}}\right)\right]-iv_{g}\eta^{2}},
\label{tt'}
\end{equation}
where $\Delta_f=E-\delta+i\gamma_{f}$, and $\gamma_{e(f)}$ is the spontaneous emission of the excited (metastable) state $\left|e(f)\right\rangle$. The corresponding transmission rate is plotted in Fig.~8. Compared with the ideal situation without relaxation, it is found that the valley becomes wider and shallower by comparing Fig.~\ref{f7}(d),(e) and Fig.~\ref{f8}(a),(b).

In the previous discussion, we have only considered the effect of spontaneous emission on the transmission rate. It should be noted that, in the presence of the natural relaxation, the conservation relation $|t_{1(2)}|^2+|r_{1(2)}|^2=1$ does not hold on any more. Therefore, it is beneficial to discuss the reflection rate. Based on Eqs.~(\ref{eq:6}) and (\ref{eq:15}), and the Sch\"{o}dinger equation, the reflection amplitudes of two- and three-level giant atom are expressed by

\begin{align}
r_{1}&=\frac{2f^{2}\left[1+\cos(kx_{0})\right]e^{ikx_{0}}}{i\left(E-\omega_{e}+i\gamma_{e}
\right)v_{g}-2f^{2}\left(1+e^{ikx_{0}}\right)},\\
r_{2}&=\frac{2f^{2}\Delta_{f}\left[1+\cos\left(kx_{0}\right)\right]e^{ikx_{0}}}
{\Delta_{f}\left[i\left(E-\omega_{e}+i\gamma_{e}\right)v_{g}-2f^{2}
\left(1+e^{ikx_{0}}\right)\right]-iv_{g}\eta^{2}}.
\end{align}}

\begin{figure}
\includegraphics[width=1\columnwidth]{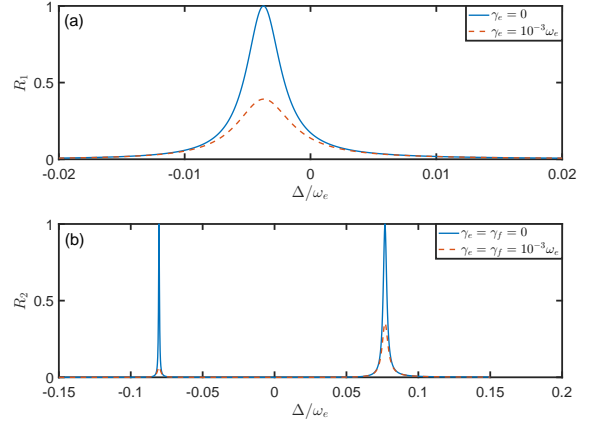}
\caption{(a) The reflection rate $R_1$ as a function of $\Delta$. (b) The reflection rate $R_2$ as a function of $\Delta$. The parameters are set as $\omega_e=3$\,GHz, $v_g=3\times10^8$\,m/s, $\omega_d=0.3\omega_e$, $\omega_f=0.7\omega_e$,$f/\sqrt{v_{g}\omega_{e}}=0.05$, $\eta=5f=0.08\omega_e$, $\gamma_e=\gamma_f=10^{-3}\omega_{e}$.}
\label{f9}
\end{figure}

In Fig.~\ref{f9}, we plot the reflection rates $R_{1}=\left|r_{1}\right|^{2}$ and $R_{2}=\left|r_{2}\right|^{2}$ as functions of $\Delta$, respectively. Comparing the results with and without spontaneous emission, it shows that the spontaneous emission breaks the complete reflection and makes the reflection peaks become wider. However, the locations of the peaks relative to the transmission valleys remain unchanged.

\section{conclusion and remarks}
\label{conclusion}
In this paper, we have studied the single-photon transmission in a one-dimensional linear waveguide system coupled with two-level or three-level giant atom.
{Generally, in the giant-atom regime, the backward and forward photons propagating in the waveguide can lead to an interference effect. Thus, for the case of two-level giant atom, we have shown that the complete reflection occurs, but not for the case that the incident photon is exactly resonant to the atom.}
In other words, the natural interference leads to an effective small but nonzero frequency shift, and the shift can be approximately regarded as a sinusoidal function of the atomic size.
{For the driven three-level giant atom, we obtain an ATS line shape, and the transparency window can be controlled by phase modulation in terms of the size of the giant atom.
In addition, the location and width of the transmission valleys are tunable by adjusting the atomic size.}

{Throughout the paper, we have considered the tunability from the viewpoint of the changing coupling distance, i.e., the giant atomic size.
Considering the superconducting transmission line as the linear waveguide, this distance can be tuned by the coupled inductance or capacitance~\cite{bk2020,AM2020}, to adjust the propagating phase $kx_0$.
Alternatively, one also can also the frequency of the atom, so that the wave vector $k$ of the resonant mode in the waveguide will be shifted.
As a result, the propagating phase will be changed, and produce the similar single-photon scattering spectrum.
However, the period on $x_0$, which is relative to $k$ [as shown in Eqs.~(\ref{eq:9},\ref{tt},\ref{rr})] will be correspondingly modulated.}

{Beyond the specific model here, our work demonstrates how to use the interplay between (among) the different physical processes to modify the photon transmission in the waveguide.
In the giant atom scenario, one of the processes is provided by interference effect during the photon propagation, while the other arises from the inner energy-level coupling, which is induced by, for example, the classical driving.
Motivating by the interplay mechanism, we hope that the giant atom can be useful in designing photon or phonon based quantum device, which goes beyond the conventional small atom setup.}

\begin{acknowledgments}
This work is supported supported by National Key R$\&$D Program of China (No. 2021YFE0193500), by  National Natural Science Foundation of China (Grant No. 11875011, No. 12047566) and the Fundamental Research Funds for the Central Universities (Grant No. 2412019FZ045).
\end{acknowledgments}
\appendix
\addcontentsline{toc}{section}{Appendices}\markboth{APPENDICES}{}
\begin{subappendices}
\section{Two small atoms}
\label{A1}

In this appendix, we give the results when the giant atom is replace by two small atoms separated by $x_0$.  Similar to the giant atom setup, we also divide the Hamiltonian of the system into three parts, i.e., $H_{3}=H_{e}+H_{\omega}+V_{3}$. The first part $H_{e}$ of the two atoms is

\begin{equation}
\label{Aeq:1}
H_{e}=\omega_{e}(\left|e\right\rangle _{1}\left\langle e\right|+\left|e\right\rangle _{2}\left\langle e\right|),
\end{equation}
where $\omega_e$ is the transition frequency between the ground state $\left|g\right\rangle $ and the excited state $\left|e\right\rangle $.

For the third part of the Hamiltonian $H_{3}$, we describe the interaction between the two small atoms and the waveguide. Within the rotating wave approximation, it can be expressed as

\begin{equation}
\begin{split}
\label{Aeq:2}
V_{3}=&f\int dx\delta\left(x\right)\left[\sigma_{1}^{+}C_{R}\left(x\right)+\sigma_{1}^{+}C_{L}
\left(x\right)+{\rm H.c.}\right]\\
&+f\int dx\delta\left(x-x_{0}\right)\left[\sigma_{2}^{+}C_{R}\left(x\right)+\sigma_{2}^{+}
C_{L}\left(x\right)+{\rm H.c.}\right].
\end{split}
\end{equation}
where $f$ is the coupling strength between the waveguide and two small atoms. The Dirac-$\text{\ensuremath{\delta}}$ function in the Hamiltonian $V_{3}$ indicates that the two small atoms are located at $x=0$ and $x=x_{0}$, respectively, and interact with the linear waveguide at these two points.

The eigenstate in the single-excitation subspace can be written as

\begin{equation}
\begin{split}
\left|E\right\rangle =&\int dx\left[\phi_{R_3}\left(x\right)C_{R}^{\dagger}\left(x\right)+\phi_{L_3}\left(x\right)C_{L}^{\dagger}\left(x\right)\right]\left|G\right\rangle \\ &+u_{1}\sigma_{1}^{+}\left|G\right\rangle +u_{2}\sigma_{2}^{+}\left|G\right\rangle,
\end{split}
\end{equation}
where $|G\rangle$ represents that the waveguide is in the vacuum states, while two atoms are in the ground state $|g\rangle$. $\phi_{R_3}\left(x\right)$ and $\phi_{L_3}\left(x\right)$ are single-photon wave functions of the right-going and left-going modes in the waveguide respectively. $u_{1}$ and $u_{2}$ are the excitation amplitudes of the two atoms, respectively. Solving the stationary Sch\"{o}dinger equation {$H_{3}\left|E\right\rangle =E\left|E\right\rangle $}, the amplitudes equation can be obtained as
\begin{subequations}
\label{Aeq:3}
\begin{eqnarray}
-iv_{g}\frac{d}{dx}\phi_{R_3}\left(x\right)+f\left[\delta\left(x\right)u_{1}+\delta\left(x-x_{0}\right)u_{2}\right]&=&E\phi_{R_3}\left(x\right),\nonumber \\
\\
iv_{g}\frac{d}{dx}\phi_{L_3}\left(x\right)+f\left[\delta\left(x\right)u_{1}+\delta\left(x-x_{0}\right)u_{2}\right]&=&E\phi_{L_3}\left(x\right),\nonumber \\
\\
\omega_{e} u_{1}+f\left[\phi_{R_3}\left(0\right)+\phi_{L_3}\left(0\right)\right]&=&Eu_{1}, \\
\omega_{e} u_{2}+f\left[\phi_{R_3}\left(x_{0}\right)+\phi_{L_3}\left(x_{0}\right)\right]&=&Eu_{2}.
\end{eqnarray}
\end{subequations}

Next, we consider the scattering behavior when a single photon with wave vector $k$ is incident from the left side of the waveguide. Similar to the discussion in two-level giant-atom setup, the wave function $\phi_{R_3}\left(x\right)$ and $\phi_{L_3}\left(x\right)$ can be expressed as

\begin{equation}
\label{Aeq:4}
\begin{split}
\phi_{R_3}\left(x\right)=&e^{ikx}\{ \theta\left(-x\right)+A_3\left[\theta\left(x\right)-\theta\left(x-x_{0}\right)\right]\\&
+t_3\theta\left(x-x_{0}\right)\},
\end{split}
\end{equation}
\begin{equation}
\label{Aeq:5}
\phi_{L_3}\left(x\right)=e^{-ikx}\left\{ r_3\theta\left(-x\right)+B_3\left[\theta\left(x\right)-\theta\left(x-x_{0}\right)\right]\right\},
\end{equation}

Then, we substitute Eqs.(\ref{Aeq:4}) and (\ref{Aeq:5}) into Eqs.(\ref{Aeq:3}), it yields the dispersion relation $E=v_{g}k$. Furthermore the transmission rate $T_3=\left|t_3\right|^{2}$ can be obtained as

\begin{equation}
T_3=\frac{\left(\Delta^{2}v_{g}\right)^{2}}{\left(\Delta^{2}v_{g}-\frac{2f^{4}}{v_{g}}\sin^{2}\left(kx_{0}\right)\right)^{2}+\left(2f^{2}\Delta+\frac{f^{4}}{v_{g}}\sin\left(2kx_{0}\right)\right)^{2}}.
\end{equation}
where $\Delta=E-\omega_e$ is the detuning between the atom and the propagating photon in the waveguide. It clearly shows that, the incident photon will completely reflected when $\Delta=0$ which is different from that of a single giant atom, as discussed in the main text.

\end{subappendices}

\end{document}